# Fast interrogation wavelength tuning for all-optical photoacoustic imaging

Jérémy Saucourt,[1] Antonin Moreau,[1] Julien Lumeau,[1] Hervé Rigneault,[1] and Thomas Chaigne[1,*]

[1] *Aix Marseille Univ, CNRS, Centrale Marseille, Institut Fresnel, Marseille, France*
*\* thomas.chaigne@fresnel.fr*

**Abstract:** Optical detection of ultrasound for photoacoustic imaging provides large bandwidth and high sensitivity at high acoustic frequencies. Higher spatial resolutions can therefore be achieved using Fabry-Pérot cavity sensors, as compared to conventional piezoelectric detection. However, fabrication constraints during deposition of the sensing polymer layer require a precise control of the interrogation beam wavelength to provide optimal sensitivity. This is commonly achieved by employing slowly tunable narrowband lasers as interrogation sources, hence limiting the acquisition speed. We propose instead to use a broadband source and a fast tunable acousto-optic filter to adjust the interrogation wavelength at each pixel within a few microseconds. We demonstrate the validity of this approach by performing photoacoustic imaging with a highly inhomogeneous Fabry-Pérot sensor.

## 1. Introduction

Over the past few decades, photoacoustic tomography has been developed to image objects embedded deep inside scattering biological tissue with optical absorption contrast. This technique relies on the ultrasound waves emitted upon the absorption of transient illumination [1]. Since these pressure waves are only weakly scattered when propagating through soft tissue, the acoustic field can be detected at the tissue surface and the optically-absorbing structures can then be reconstructed with acoustic resolution [2].

Ultrasound can be either measured with piezoelectric sensors, capacitive micromachined ultrasound transducers (CMUT), or optical sensors [3–5]. The latter usually exhibit a larger bandwidth, as well as a better sensitivity at high frequencies. Moreover, their transparency enables to illuminate the sample and measure the emitted acoustic field from the same aperture. This prevents cumbersome configurations that are typically encountered with other conventional detection schemes.

The most widespread optical detection technique for photoacoustic imaging is based on a Fabry-Pérot (FP) cavity, with a polymer spacer inserted between two mirrors [6,7]. These two mirrors are transparent for the transient illumination, but semi-reflective on a distant wavelength range. Due to interferences within the cavity, the reflection spectrum of the FP sensor within this range exhibits dips at particular wavelengths that depend on the thickness of the spacer in between the two mirrors. When placing the FP sensor in acoustic contact with the tissue of interest, the outgoing ultrasound waves propagate through the polymer spacer and modulate its thickness. This locally modulates the reflection spectrum of the FP cavity, which can be probed by focusing a so-called interrogation beam at the surface of the FP sensor and measuring the reflected light with a fast photodiode. The local ultrasound signal can then be measured by setting the interrogation wavelength at the highest slope of the rising edge of one of the dips [7]. To reconstruct the image, i.e. the initial pressure source following the optical nanosecond pulse absorption, this measurement has to be repeated while scanning the interrogation beam across the sensor area, providing a time-dependent two-dimensional ultrasound field.

The performances of this technique are affected by fabrication constraints of the FP sensor, in particular the thickness homogeneity of the polymer layer. For a total thickness L of a few tens of micrometers, fluctuations $\Delta L$ of this thickness around a few tens of nanometers (corresponding to a typical surface quality of $\lambda/10$ for mid-range optical components) would results in a spectral shift of the reflection spectrum $\Delta\lambda$ up to a few nanometers, following the equation $\frac{\Delta\lambda}{\lambda} = \frac{\Delta L}{L}$.

The interrogation wavelength $\lambda$ must then be adjusted pixel-to-pixel to compensate for these thickness fluctuations and to keep it as close as possible to the highest sensitivity range. Typical continuous-wave interrogation sources have a limited tuning speed of the order of 1-10 nm/s. Faster interrogation sources such as vertical cavity surface emitting laser (VCSEL) have been introduced, but only achieve single pixel tuning time of a few milliseconds [8]. This significantly limits the final frame rate, or conversely sets stringent constraints on the fabrication tolerance of the polymer spacer, increasing the complexity of coating steps and associated costs.

As a result, most FP sensors are nowadays fabricated using chemical vapor deposition (CVD) of Parylene C, which provides a suitable finishing state [8,9], hence reducing the number of necessary wavelength tuning steps. However, this fabrication is in practice restricted to a few polymer types, leaving out some polymers that could better fit in terms of acoustic properties or for further patterning steps. This step also requires specific expensive deposition equipment, which are usually only available in high-end technical platforms. Consequently, Parylene C-based FP sensors cannot be treated as consumables, limiting their use as surgical implants in animals for instance.

Other polymer coating techniques like spin coating are usually available or affordable by most labs, and can be used with a variety of polymers, as long as they can be dissolved [10]. However, although many parameters can be tuned to control the overall thickness of the deposited film, its homogeneity across the entire substrate is not as precise as what can be achieved with CVD deposition [9]. Strategies to correct these fluctuations by further tuning the local optical refractive index (reversibly or not) have been proposed, either using electro- or photo-sensitive polymers [11]. This nonetheless requires additional processing steps and more complex sensor design and fabrication.

The combination of both affordable FP sensors and high imaging rates therefore requires a fast tuning of the interrogation source wavelength. Here we introduce a new scheme for this purpose: instead of shifting the wavelength of a narrowband source, we use a broadband source and select the optimal interrogation wavelength at any pixel of the FP sensor with a narrow, fast-tunable filter. We implement this approach using a broadband amplified spontaneous emission (ASE) source and an acousto-optic tunable filter (AOTF). We demonstrate its performances on a FP sensor based on a Parylene C spacer (deposited by CVD), and on a second one based on a SU-8 photoresist (deposited by spin-coating). Additionally, we show that fine tuning of the interrogation wavelength is crucial to detect the high frequency content of the photoacoustic signal, and thus provide high resolution images.

## 2. Experimental setup, calibration, and acquisition

The tri-dimensional image of the optical absorption distribution is formed by reconstructing the initial pressure rise, following the nanosecond pulsed illumination of the medium. This requires measuring the time-varying ultrasonic field over a large area at the surface of the medium. As depicted in **Fig. 1**, the imaging system can be split in three main parts: an interrogation source, a raster-scanning microscope, and a detection system.

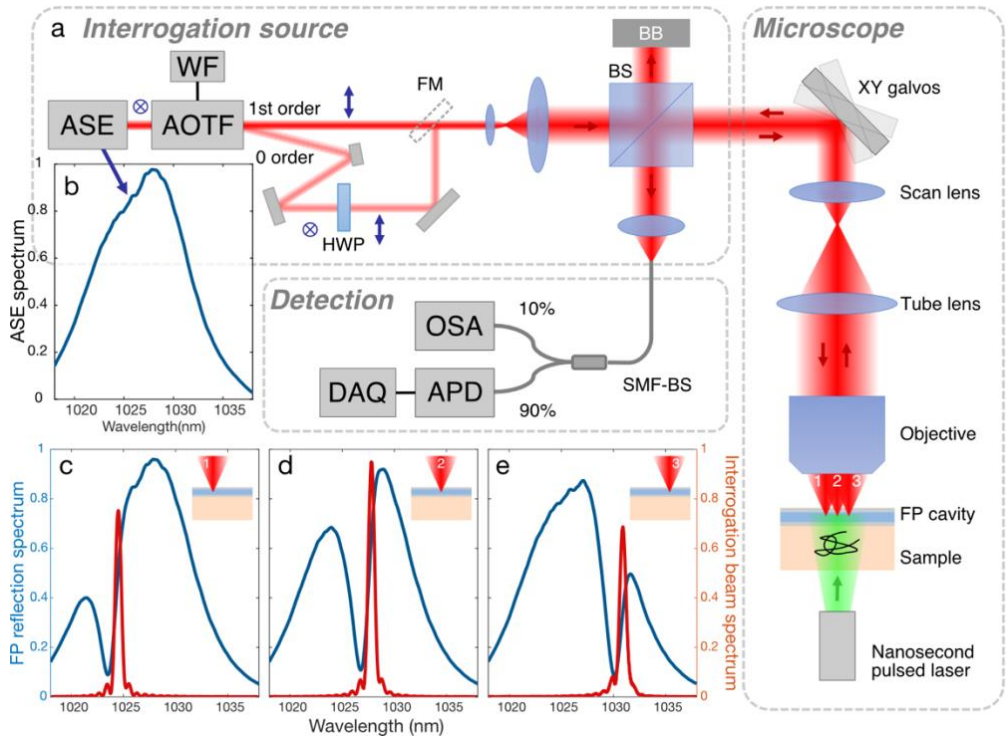

**Fig. 1. Setup and calibration method.** a) Schematic of the experimental setup – Interrogation source: ASE: amplified spontaneous emission source; AOTF: acousto-optic tunable filter; WF: waveform generator; HWP: half waveplate; FM: flip mirror; BS: beam splitter (50:50); BB: beam block. Detection: SMF-BS: single mode fiber splitter (10:90); OSA: optical spectrum analyzer; APD: avalanche photodiode; DAQ: data acquisition system. b) Spectrum of the ASE source, measured in reflection using a mirror in place of the FP cavity. c-d-e) Blue: reflection spectra of the FP cavity for three different positions of the interrogation beam (1, 2, 3, see insets), when illuminating with the full ASE spectrum (AOTF off, 0 order path); red: interrogation beam spectra when selecting the wavelength corresponding to the highest slope of the FP spectra at each position (AOTF on, 1$^{st}$ order path).

The interrogation source uses a linearly-polarized broadband amplified spontaneous emission (ASE) source (VASP-Yb-B-150-0-C, Connet), with a -3dB bandwidth of 10 nm, centered around 1027 nm (as shown in **Fig. 1.b**). The beam is spectrally filtered by an acousto-optic tunable filter (AOTF) (TF950-500-1-2-GH96, G&H), driven by a waveform generator (Wavepond DAX22000-8M, Chase Scientific). This acousto-optic filter is based on a birefringent crystal whose ordinary and extra-ordinary optical refractive indices can be modulated by an acoustic wave propagating through the crystal. A sinusoidal modulation of these indices sets phase matching conditions resulting in the diffraction of a narrow band of the incident light spectrum. The combined properties of this filter and generator allow to switch between two different RF frequencies and thus two different diffracted optical wavelengths in just 15μs, independently of the extent of the wavelength shift. The spectral linewidth of the filtered diffracted beam is about 0.5nm. Using a flip-mounted mirror, either the undiffracted beam or the first diffraction order can then be sent into a custom-made laser-scanning microscope (XY galvanometric scanners: 6215HSM40B, Cambridge Technology; scan lens: LSM04-BB, Thorlabs; tube lens: AC508-200-B, Thorlabs; objective: 4X Plan Fluorite, NA 0.13, Nikon). The light reflected by the FP cavity is focused into a single-mode fiber, as mode filtering has been shown to improve acoustic detection sensitivity [12]. This fiber is actually the input side of a 90:10 fiber splitter (TW1064R2A1B, Thorlabs), which splits the reflected light in two parts: the 10% arm is sent to an optical spectrum analyzer (OSA) (AQ6370D,

Yokogawa), and the 90% part to an avalanche photodiode (APD430A/M, Thorlabs). The signal is digitized at 500MS/s using a fast acquisition card (RMX-165-020, Gage).

The FP sensors are based on a 1 mm thick glass substrate with a 25 mm diameter (B270, Schott). An anti-reflection coating is deposited on the face opposed to the FP cavity to avoid interferences between reflections from both sides of the substrate. The dielectric mirrors forming the cavity are made of 9 alternating quarter-wave layers of high index and low index materials (respectively at 1000 nm: ZnS, n = 2.36; YF3, n = 1.49). They were fabricated by electron-beam deposition (Bühler SYRUSpro 710). Two kinds of polymer spacers are used: a 3.1 µm layer of Parylene C deposited by CVD (Labcoter, SCS), or a 10.4 µm layer of SU-8 (3010, Kayaku Microchem) deposited by spin-coating (WS-400, Laurell). For the Parylene C spacer, an additional layer of low index material is deposited on the polymer spacer to finely tune the optical thickness of the spacer to a half-wave layer at 1026 nm. This step ensures that the reflection spectrum of the FP cavity exhibits a dip in the center of the ASE source spectrum.

Absorbing objects (**Fig. 2-3(a-f)-4**: 20 µm black nylon wire (NYL02DS, Vetsuture); **Fig. 3(g-j)**: black polyethylene 10-20 µm beads (Cospheric)) are embedded in a 1.2% (mass ratio) agarose gel block. This phantom sample is immersed in de-ionized water and placed under the FP sensor. A nanosecond pulsed laser (wavelength: 515 nm, repetition rate: 2 kHz, pulse duration: 1.2 ns, mean power: 400 mW, Flare NX, Coherent) illuminates the sample from below with a beam diameter of 5 mm, yielding a fluence rate of 20 mW/mm$^2$, or equivalently a fluence of 10µJ/mm$^2$ per pulse at 2 kHz.

The FP cavity needs to be calibrated before the acquisition of the photoacoustic signals. The AOTF is turned off and the flip mirror is placed so that the full ASE spectrum illuminates the FP cavity. The focused interrogation beam (5 µm full width at half maximum) is then scanned over a 2 mm field-of-view (FOV) with a 20 µm step. For each position, the reflection spectrum of the FP cavity is measured with the OSA, and the wavelength corresponding to the maximum derivative within the appropriate linear range of the spectrum is computed and stored. This calibration procedure yields a map of optimal wavelengths (and corresponding RF frequencies to be sent to the AOTF) that should be used at each pixel to get the highest sensitivity (see **Fig. 2.h**).

To acquire PA signals, the FP sensor is scanned using the 1$^{st}$ diffraction order beam of the AOTF across the same 2 mm FOV with a 20 µm step. The waveform generator drives the AOTF so that the optimal interrogation wavelength is used at each pixel, with a wavelength tuning time of only 15 µs. The acquisition of each photoacoustic signal is triggered by the emission of a nanosecond pulse. 500 signals are averaged for each pixel to provide a high signal-to-noise ratio. The resulting signal is normalized by the intensity of the ASE source at the local optimal wavelength, and by the local slope. This compensates for the uneven emission spectrum of the ASE source and for the local optical properties of the FP sensor. The signals are then filtered between 10 and 120 MHz (or other bands specified further) using a digital first order Butterworth bandpass filter, and the photoacoustic images are reconstructed using a custom delay-and-sum beamforming algorithm [2]. Finally, the modulus of the Hilbert transform of this reconstructed volume is computed and maximum intensity projection are used for display in the following figures.

## 3. Results

We first test our approach using the Parylene C-based sensor. **Fig. 2**.h shows the calibration map of wavelengths maximizing the optical sensitivity at each pixel. The colormap has been adapted to reflect the distribution of these wavelengths (**Fig. 2**.i), with black pixels depicting the invalid virtual elements for which no reflection dip could be distinguished within the available ASE spectrum. These pixels are ignored in the acquisition and reconstruction steps. Despite the good thickness homogeneity yielded by the CVD process, we observe that the optimal wavelengths are still spread over a few nanometers.

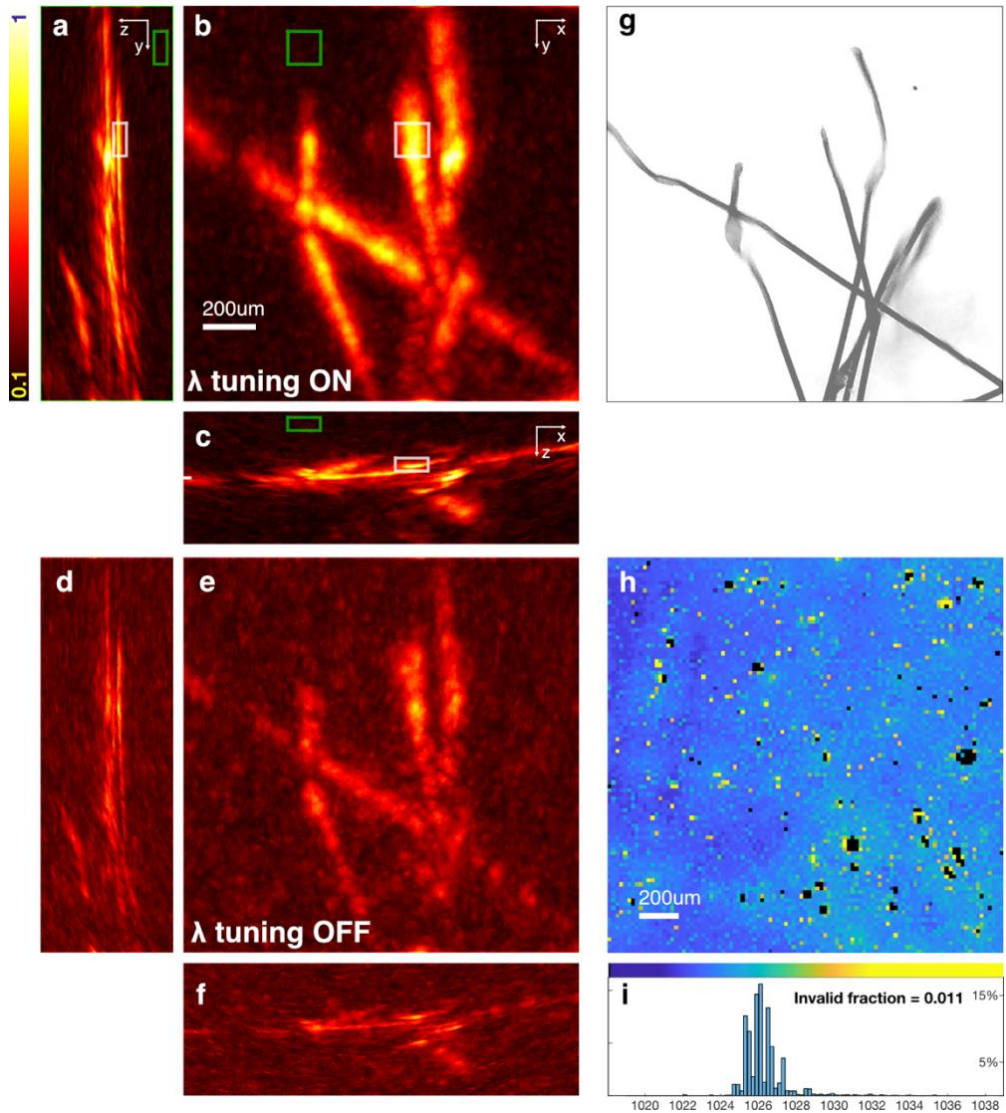

**Fig. 2. Photoacoustic imaging using fast wavelength tuning for maximum sensitivity.** a-b-c) Maximum intensity projections (MIP) of the reconstructed volume when adjusting the interrogation wavelength to the maximum sensitivity of the FP sensor at each pixel. Squares in (b) delineate the regions of interest that are used to compute the signal (white) to background (green) ratio. The white middle left mark in (c) depicts the position z = 2.2 mm from the FP sensor. d-e-f) MIP of the reconstructed volume when using only the mean wavelength of the calibration map for all pixels of the FP sensor. All images in (a-f) are normalized to the maximum value of the reconstructed volume with wavelength control (a-c). They are represented with the same colormap, depicted by the color bar on the left. g) Ground truth picture of the phantom sample (20 µm black nylon wire). h) Calibration map of the wavelengths corresponding to the highest slope of the reflection spectrum at each pixel of the FP sensor. Black pixels depict positions at which no reflection dip was detected, and thus are skipped during the acquisition and reconstruction. The colormap is adjusted to the following histogram. i) Distribution of optimal wavelengths across the FP sensor (percentage of total number of pixels). The invalid fraction is the ratio between the number of unused black pixels in (h) and the total number of pixels scanned on the FP sensor. Scale bars: 200 µm.

To highlight the relevance of our wavelength tuning strategy, we image a phantom sample with black 20 µm wires (**Fig. 2**.g) either using the full calibration of the FP sensor (**Fig. 2**.a-c), or using the mean wavelength of this calibration (**Fig. 2**.d-f). The signal-to-background ratio (SBR) (computed as the ratio between the mean values within the regions of interest depicted with white and green squares in **Fig. 2**.b) is thus strongly enhanced when using the full calibration, raising from 2.58 to 4.62.

We further investigate the effect of this fine wavelength tuning on the high frequency content of the detected ultrasound signals. In **Fig. 3**, we reconstruct images of the black wires after digitally band-pass filtering the PA signals either between 10 MHz and 120 MHz (**Fig. 3**.a,b), 30 MHz and 120 MHz (**Fig. 3**.c,d), or 50 MHz and 120 MHz (**Fig. 3**. e,f). We observe that precise tuning of the interrogation wavelength is critical to provide the highest sensitivity for large ultrasonic frequencies, and thus the best possible resolution. We also demonstrate this in **Fig. 3**.g-l by imaging a collection of 10-20 µm black beads, and reconstructing the image after filtering the signals on the same three frequency bands. The small beads remain visible in the 50-120 MHz band when compensating for the cavity spacer inhomogeneities but vanish otherwise.

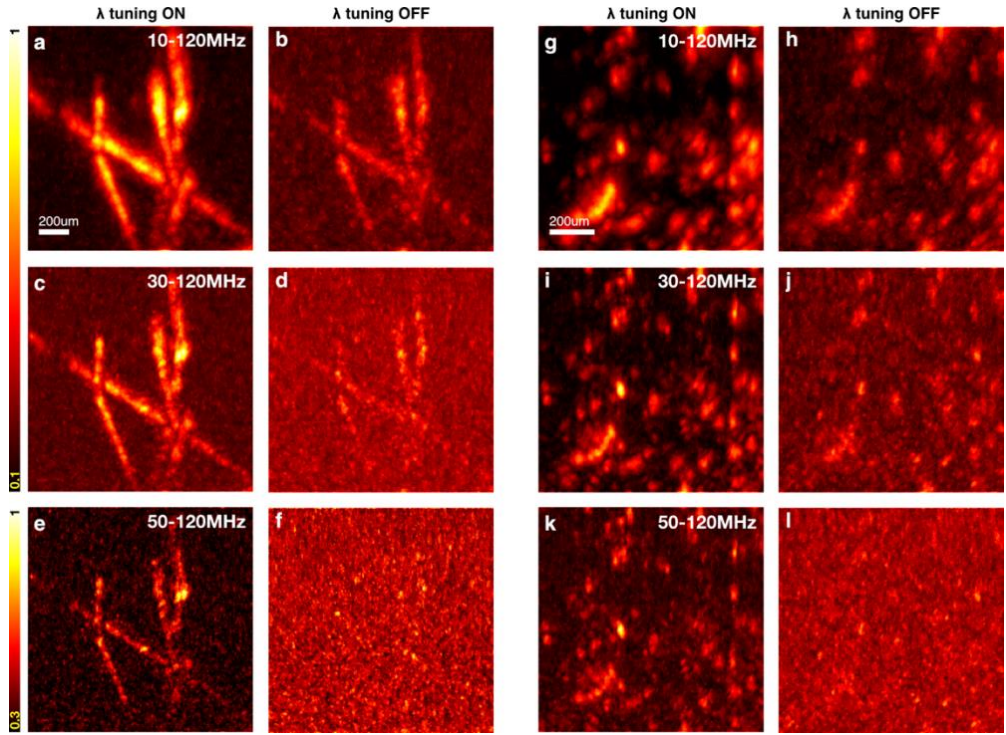

**Fig. 3. High ultrasound frequencies can only be detected when using optimal wavelengths.**
a-d) Phantom sample: 20 µm black nylon wires - MIP (along z axis) of the reconstructed volume after filtering the PA signals: between 10 MHz and 120 MHz, a) with optimal wavelength tuning and b) with mean wavelength; between 30 MHz and 120 MHz, c) with optimal wavelength tuning and d) with mean wavelength; between 50 MHz and 120 MHz, e) with optimal wavelength tuning and f) with mean wavelength. g-l) Phantom sample: 10-20 µm black beads, located around z = 1.5 mm from the FP sensor. - MIP (along z axis) of the reconstructed volume after filtering the PA signals between 10 MHz and 120 MHz, g) with wavelength tuning and h) with mean wavelength; between 30 MHz and 120 MHz, i) with wavelength tuning and j) with mean wavelength; between 50 MHz and 120 MHz, k) with wavelength tuning and l) with mean wavelength. All pairs of images are normalized to the maximum value of the left one (reconstructed volume with wavelength tuning). The color bars on the left are valid for all images within the same line. Scale bars: 200 µm.

To further demonstrate the validity of our approach, we image again a phantom sample containing 20 µm black wires, now with a FP cavity containing a polymer spacer made of spin-coated SU-8 resist. The very poor thickness homogeneity of this layer can be observed in **Fig. 4**.g. As opposed to the Parylene C-based FP cavity, the optimal wavelengths are equally spread over the entire available ASE spectrum (**Fig. 4**.h), and almost half of the pixels do not exhibit a reflection dip within this band (depicted in black in **Fig. 4**.g). Despite these adverse conditions, we are still able to perform PA imaging using this FP sensor, as **Fig. 4** demonstrates. When optimally tuning the wavelength, the black wires distinctly appear (**Fig. 4**.a-c), although the various regions of the image do not all exhibit the same reconstruction quality. Due to the wide spreading of the optimal wavelengths, the objects cannot be reconstructed when considering the sole average wavelength, as can be seen in **Fig. 4**.d-f.

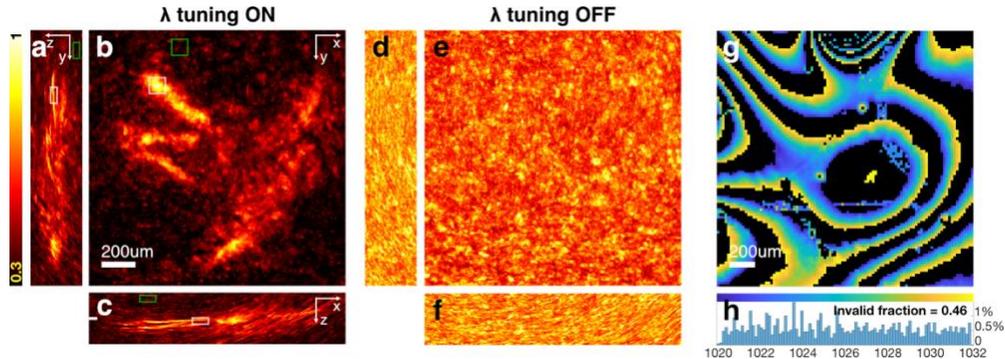

**Fig. 4. Photoacoustic imaging using a highly inhomogeneous FP sensor.** MIP of the reconstructed volume after filtering the PA signals between 10 and 120 MHz, a-c) with optimal wavelength tuning, or d-f) with mean wavelength. The white middle left mark in (c) depicts the position z = 2.1 mm from the FP sensor. g) Calibration map showing interrogation wavelengths that provide maximum sensitivity at each pixel of the FP sensor. h) Distribution of optimal wavelengths across the FP sensor. Scale bars: 200 µm.

## 4. Conclusion

We presented a new method for optical interrogation of FP sensors for ultrasound field measurement. Instead of relying on a slowly tunable narrowband laser source, we use a broadband ASE source and take advantage of the short microsecond response time of an acousto-optic tunable filter. This allows us to quickly select the most adequate interrogation wavelength at each pixel of the FP sensor. The primary advantage of this approach is the ability for the interrogation beam to hop from one wavelength to another in a short amount of time, independently of their values. This provides a significant asset compared to tunable laser sources which need to scan the wavelength through the entire range separating two values. We also stress that the proposed interrogation source provides a few mW of optical power on the FP sensor, which is comparable to what is achieved with conventional telecom tunable laser sources [7,13].

One drawback of this approach is the limited spectral width of the ASE source (10 nm at -3 dB). This requires to precisely control the average thickness of the cavity spacer to ensure that there will be a dip in the reflection spectrum within this range. However, the inaccuracies of the polymer deposition can be compensated by adding low index dielectric material before the second mirror, as mentioned earlier in the description of the FP sensor fabrication.

A second limitation is the spectral width of the filtered interrogation beam, which mostly depends on the interaction length of the light with the birefringent acousto-optic crystal [14]. This length is about 1 cm for the AOTF used in this work, yielding a spectral full width at half maximum of the 1$^{st}$ diffraction order of 0.5 nm. This sets an upper limit for the slope of the reflection spectrum of the sensor, which is here about 0.5-1 nm$^{-1}$. On the contrary, tunable laser sources can have linewidth down to a few tens of MHz or less (>0.001 nm at 1500 nm), and can be potentially used with FP sensors exhibiting a higher optical sensitivity.

To address this issue and emphasize the interest of our approach, we estimate the total acquisition time with our fast tunable interrogation source (15 µs per wavelength shift), and for a typical tunable laser source (wavelength scanning speed: 1nm/s). For both interrogation sources, the same excitation repetition rate of 2 kHz is considered. In this paper, the median wavelength shifts from one pixel to the next are 0.4 nm for the Parylene C-based sensor and 1.8nm for the SU-8-based one. We first consider the same FP sensor with a given sensitivity and thus a similar signal averaging with both interrogation techniques. For the AOTF-based interrogation technique, the acquisition duration at one single pixel is dominated by the averaging time: 15 µs+500×500 µs = 250.015 ms. For the tunable laser-based interrogation, the acquisition duration is dominated by the wavelength shift: 400 ms+500×500 µs = 650 ms for the Parylene-based sensor, and 1.8 s+500×500 µs = 2.05 s for the SU-8-based sensor. As a consequence, even if this last interrogation system can easily afford a higher optical sensitivity by at least a factor 10 [15], requiring 100 times less averaging to reach the same signal-to-noise ratio, the acquisition duration is not reduced as much: 400 ms+5×500 µs = 402.5 ms for the Parylene-based sensor, and 1.8 s+5×500 µs = 1.8025 s for the SU-8-based sensor. Therefore, the approach we introduce here remains advantageous.

The wavelength shifting time of the AOTF is comparable to the minimum time per pixel set by the galvanometric scanners due to inertia. The maximum line rate is indeed about 1 kHz, yielding a pixel dwell time of 10 µs for 100 pixels per line, or equivalently a pixel rate of 100kHz. The ultimate limitation is then set by the 2 kHz repetition rate of the nanosecond excitation laser, still much lower than the 67 kHz that could be achieved with our technique. However, the noise level in our setup does not allow us yet to fully benefit from this repetition rate, as we need to average several PA signals per pixel.

Nonetheless, we showed that this technique could be used to perform photoacoustic imaging with FP sensor even when the thickness of the polymer spacer was highly inhomogeneous. Fast optical compensating of the low quality of the polymer deposition could allow the use of polymers that cannot be deposited by CVD, potentially with more adequate acoustic properties. This also opens the way towards cheaper and disposable FP cavities, which is still a strong requirement for implantable sensors [16].

**Funding.** CNRS Momentum (2018), CR-PACA (204-204181), CD-13 (213962), ANR-21-ESRE-0003 CIRCUITPHOTONICS

**Acknowledgments.** The authors thank Olivier Hector and Dr. David Moreau for technical support during thin film deposition; Prof. David Grosso and Martin O' Byrne for technical support and advice for spin-coating operations; Prof. Emmanuel Bossy and Dr. Guillaume Godefroy for their help with the 3D reconstruction algorithm, and Prof. Bossy for valuable comments on the manuscript.

**Disclosures**. The authors declare no conflicts of interest.

**Data availability.** Data underlying the results presented in this paper are not publicly available at this time but may be obtained from the authors upon reasonable request.